\documentclass{article}
\usepackage{spconf,amsmath,graphicx}
\usepackage{epstopdf}
\usepackage{algorithm} 
\usepackage{algorithmic} 
\usepackage{multirow} 

\usepackage{amsmath}
\usepackage{amsthm}

\title{DEEP LEARNING BASED ON orthogonal approximate message passing FOR CP-FREE OFDM}
\name{Jing Zhang$^\ast$, Hengtao He$^\ast$, Chao-Kai Wen$^\dagger$, Shi Jin$^\ast$, Geoffrey Ye Li$^\ddag$\thanks{This work was supported in part by the National Science Foundation (NSFC) for Distinguished Young Scholars of China with Grant 61625106, and in part by the NSFC under Grant 61531011. The work of C.-K. Wen was supported by the Ministry of Science and Technology of Taiwan under Grants MOST 107-2221-E-110-026 and the ITRI in Hsinchu, Taiwan. Email: jingzhang@seu.edu.cn}}
\address{$^\ast$ National Mobile Communications Research Laboratory, Southeast University, China\\$^\dagger$ Institute of Communications Engineering, National Sun Yat-sen University, Taiwan\\$^\ddag$ School of Electrical and Computer Engineering, Georgia Institute of Technology, USA}
\begin{document}
%
\maketitle
\begin{abstract}
Channel estimation and signal detection are very challenging for an orthogonal frequency division multiplexing (OFDM) system without cyclic prefix (CP). In this article, deep learning based on orthogonal approximate message passing (DL-OAMP) is used to address these problems. The DL-OAMP receiver includes a channel estimation neural network (CE-Net) and a signal detection neural network based on OAMP, called OAMP-Net. The CE-Net is initialized by the least square channel estimation algorithm and refined by minimum mean-squared error (MMSE) neural network. The OAMP-Net is established by unfolding the iterative OAMP algorithm and adding some trainable parameters to improve the detection performance. The DL-OAMP receiver is with low complexity and can estimate time-varying channels with only a single training. Simulation results demonstrate that the bit-error rate (BER) of the proposed scheme is lower than those of competitive algorithms for high-order modulation.
\end{abstract}
\section{Introduction}
\label{sec:intro}
Orthogonal frequency division multiplexing (OFDM) has been widely applied in digital audio and video broadcasting, WLAN, WiMax, and 4G LTE systems. In many environments, the transmitted signal experiences multiple paths with different delays, which causes intersymbol interference (ISI) and intercarrier interference (ICI) at the receiver [1]. The ISI and ICI can be considerably mitigated by inserting sufficient cyclic prefixes (CPs) at a transmitter and properly choosing OFDM symbol duration [2, 3]. However, adding a CP in OFDM consumes additional bandwidth and consequently reduces spectrum efficiency, transmission rate, and energy efficiency, which motivates CP-free OFDM. Channel estimation and signal detection in a CP-free OFDM system face many challenges. Traditional solutions, including residual ISI cancellation (RISIC) [4-6] and symbol cyclic-shift equalization (SCSE) [7], have many limitations, such as sensitivity to channel delay spread and feedback delay.

Deep learning (DL) has been recently introduced to physical layer communications [8-12].
In [11], a fully connected deep neural network (FC-DNN) has been used to replace both the channel estimation and signal detection modules in the traditional OFDM system, which works well even for a CP-free OFDM system with QPSK modulation. A model-driven DL method, named ComNet [12], has been developed to improve the performance of the OFDM receiver, especially with high-order modulation. In general, the model-driven DL methods construct the network topology based on some known physical mechanism and professional knowledge, which can significantly reduce the required training data and time [13]. However, for CP-free OFDM, the recurrent neural network can be applied in ComNet, which can significantly improve the bit-error rate (BER) of signal detection but is with high complexity at the same time. The trainable iterative soft thresholding algorithm (TISTA) in \cite{TISTA}, which is also a model-driven method, unfolds the orthogonal approximate message passing (OAMP) algorithm and trains some variables by DL to solve the problem of sparse signal recovery.

In this article, a model-driven DL based on OAMP (DL-OAMP) for CP-free OFDM is proposed. The DL-OAMP receiver includes channel estimation neural network (CE-Net) and OAMP detection neural network (OAMP-Net). Compared with that in ComNet [12], the channel estimation module remains, but the detection structure is completely transformed. The detection part is replaced by an OAMP-Net, which is inspired by [14] and combines the OAMP algorithm and DL by introducing a few trainable parameters. Furthermore, the DL-OAMP receiver has lower complexity than ComNet and is adaptive to time-varying channels. Simulation results reveal that the proposed DL-OAMP receiver offers remarkable performance and attains lower BER than the existing algorithms, especially with high-order modulation.

\section{System Model}
\label{sec:format}
Figure 1 shows the DL-OAMP receiver. It includes CE-Net for channel estimation and OAMP-Net based signal detection. The input of CE-Net, $\mathbf{y}_{p}$, is the received signal corresponding to the pilot symbol. It first goes through a serial-to-parallel (S/P) converter and then performs FFT. The initialization block, LS\_init, is equipped with least square (LS) channel estimation to get ${{\mathbf{\hat{H}}}_{\text{LS}}}$ which initializes MMSE\_Net to generate accurate channel estimation, $\mathbf{\hat{H}}$. The signal detection component consists of OAMP-Net and demodulation module, which is completely distinct from the detection subnet in ComNet [12]. The OAMP-Net uses the iterative OAMP algorithm in [15] as the initialization and then adds some adjustable parameters to improve the detection performance.
\begin{figure}[!h]
\centering
\includegraphics[width=2.8in]{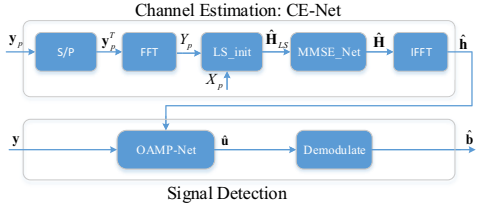}
\caption{DL-OAMP receiver for CP-free OFDM system.}
\label{fig_sim}
\end{figure}

For a CP-free OFDM system with $N$ subcarriers, its received signal vector $\mathbf{y}=[{{y}_{1}},{{y}_{2}},\ldots ,{{y}_{N}}]^T$ can be expressed as
\begin{align}
   \mathbf{y}
  &=\mathbf{Hq}-\mathbf{Aq}+\mathbf{A}{{\mathbf{q}}_{i-1}}+\mathbf{w} \\\nonumber
 &=\mathbf{H}{{\mathbf{F}}^{H}}\mathbf{u}-\mathbf{A}{{\mathbf{F}}^{H}}\mathbf{u}+\mathbf{A}{{\mathbf{q}}_{i-1}}+\mathbf{w},
\end{align}
where $\mathbf{w}$ is additive white Gaussian noise vector, $\mathbf{u}$ is the transmit symbol vectors, ${\mathbf{q}}$ and ${\mathbf{q}}_{i-1}$ represent the current and the previous OFDM signal vector, $\mathbf{F}$ is a $N\times N$ normalized FFT matrix,
\[\mathbf{H}={{\left[ \begin{matrix}
   {{h}_{0}} & 0 & \cdots  & 0 & {{h}_{L-1}} & \cdots  & {{h}_{2}} & {{h}_{1}}  \\
   {{h}_{1}} & {{h}_{0}} & 0 & \cdots  & 0 & {{h}_{L-1}} & \cdots  & {{h}_{2}}  \\
   \vdots  & \ddots  & \text{ } & \text{ } & \text{ } & \ddots  & \text{ } & \vdots   \\
   0 & \cdots  & 0 & {{h}_{L-1}} & {{h}_{L-2}} & \cdots  & {{h}_{1}} & {{h}_{0}}  \\
\end{matrix} \right]}_{N\times N}}\]
is an $N\times N$ cyclic channel matrix,
and
\[\mathbf{A}={{\left[ \begin{matrix}
   0 & \cdots  & 0 & {{h}_{L-1}} & \cdots  & \cdots  & {{h}_{1}}  \\
   0 & \cdots  & 0 & 0 & {{h}_{L-1}} & \cdots  & {{h}_{2}}  \\
   \vdots  & \cdots  & \vdots  & \ddots  & \ddots  & \ddots  & \vdots   \\
   0 & \cdots  & 0 & \ddots  & \ddots  & 0 & {{h}_{L-1}}  \\
   \vdots  & \cdots  & \vdots  & \ddots  & \ddots  & \ddots  & \vdots   \\
   0 & \cdots  & 0 & 0 & \ddots  & \cdots  & 0  \\
\end{matrix} \right]}_{N\times N}}\]
is an $N\times N$ cut-off channel matrix.

The second and third terms of (1) represent ICI and ISI, respectively. If there exists sufficient CP in the OFDM system, then $\mathbf{A=0}$ and there is no ICI or ISI.

Eq. (1) can be transformed into
\begin{align}
\nonumber
   \mathbf{y}&=\left( \mathbf{H}-\mathbf{A} \right){{\mathbf{F}}^{H}}\mathbf{u}+\mathbf{A}{{\mathbf{q}}_{i-1}}+\mathbf{w} \\
 & ={{\mathbf{H}}_{1}}{{\mathbf{F}}^{H}}\mathbf{u}+\mathbf{A}{{\mathbf{q}}_{i-1}}+\mathbf{w}
\end{align}
Denote $\mathbf{s}={{\mathbf{H}}_{1}}{{\mathbf{F}}^{H}}\mathbf{u}$ as the signal received in time domain. Thus, the signal-to-noise ratio at the receiver can be expressed as
$\text{SNR}=10{{\log }_{10}}({\mathbf{\bar{s}}}/{\sigma _{w }^{2}}),$
where $\mathbf{\bar{s}}=\mathbf{E}\{ {{\left| \mathbf{s} \right|}^{2}} \}$ and $\sigma _{w}^{2}$ represents the variance of $\mathbf{w}$.
\section{Deep Learning Based on OAMP}
\label{sec:pagestyle}
In this section, we describe CE-Net for channel estimation and OAMP-Net for signal detection in detail.
\subsection{CE-Net}
The key components for CE-Net in Fig.1 are LS\_init and MMSE\_Net. The LS channel estimation, ${{\mathbf{\hat{H}}}_{\text{LS}}}$, is obtained just by dividing $Y_{p}$ and $X_{p}$ on each subcarrier $n$ as ${{\mathbf{\hat{H}}}_{\text{LS}}(n)}={Y_{p}(n)}/{X_{p}(n)}$,
where $Y_{p}(n)$ and $X_{p}(n)$ represent the received pilot signal and transmit pilot symbol in frequency domain, respectively.

As in [16], the linear minimum mean-squared error (LMMSE) channel estimation can be obtained by
\begin{equation}\label{2}
 {{\mathbf{\hat{H}}}_{\text{LMMSE}}}={{\mathbf{W}}_{\text{LMMSE}}}{{\mathbf{\hat{H}}}_{\text{LS}}}={{\mathbf{R}}_{\mathbf{H\hat{H}_{\text{LS}}}}}{{\left( {{\mathbf{R}}_{\mathbf{HH}}}+\frac{\sigma _{\omega}^{2}}{E_s}\mathbf{I} \right)}^{-1}}\mathbf{\hat{H}_{\text{LS}}},
\end{equation}
where ${\mathbf{W}}_{\text{LMMSE}}$ denotes the $N\times N$ complex weight matrix. The corresponding real-valued form can be expressed as
\begin{equation}\label{6}
  {{\mathbf{\tilde{H}}}_{\text{LMMSE}}}={{\mathbf{\tilde{W}}}_{\text{LMMSE}}}{{\mathbf{\tilde{H}}}_{\text{LS}}},
\end{equation}
where
\[{{\mathbf{\tilde{H}}}_{\text{LMMSE}}}=\left[ \begin{matrix}
 \rm Re\{{{{\mathbf{\hat{H}}}}_{\text{LMMSE}}}\} \\
  \rm Im\{{{{\mathbf{\hat{H}}}}_{\text{LMMSE}}}\}
\end{matrix} \right],
{{\mathbf{\tilde{H}}}_{\text{LS}}}=\left[ \begin{matrix}
\rm Re\{{{{\mathbf{\hat{H}}}}_{\text{LS}}}\} \\
 \rm Im\{{{{\mathbf{\hat{H}}}}_{\text{LS}}}\} \\
\end{matrix} \right],\]
\[{{\mathbf{\tilde{W}}}_{\text{LMMSE}}}=\left[ \begin{matrix}
   \rm \rm Re\{{{\mathbf{W}}_{\text{LMMSE}}}\} & \rm Im\{{{\mathbf{W}}_{\text{LMMSE}}}\}  \\
   \rm Im\{{{\mathbf{W}}_{\text{LMMSE}}}\} & \rm Re\{{{\mathbf{W}}_{\text{LMMSE}}}\}  \\
\end{matrix} \right].\]

MMSE\_Net is a simple neural network with one input layer and one output layer. The weights are initialized by using the real-valued LMMSE channel estimation weight, ${{\mathbf{\tilde{W}}}_{\text{LMMSE}}}$. The input of MMSE\_Net is real-value  ${{\mathbf{\tilde{H}}}_{\text{LS}}}$, which has $2N$ neutrons. The number of neurons in the output layer is also $2N$ and these output neurons have no activation function.

\subsection{OAMP-Net for CP-free OFDM}
The OAMP algorithm [14, 15, 17, 18] is used for signal detection in a CP-free OFDM system. In addition, the OAMP-Net is applied to improve the performance since the pilot value of $\mathbf{q}_{i-1}$ in the previous block is known at the receiver.

The interference from the previous OFDM blocks, corresponding to the second term in (2), must be eliminated to apply the OAMP algorithm. As depicted in Fig. 1, the receiver acquires the CSI, $\mathbf{\hat{h}}$, by CE-Net. After removing residual ISI, the received signal can be expressed as
\begin{align}
   \mathbf{\hat{y}}&=\mathbf{y}-\mathbf{\hat{A}}{{{\mathbf{{q}}}}_{i-1}} \\\nonumber
 & ={{\mathbf{H}}_{1}}{{\mathbf{F}}^{H}}\mathbf{u}+\mathbf{A}{{\mathbf{q}}_{i-1}}+\mathbf{w}-\mathbf{\hat{A}}{{{\mathbf{{q}}}}_{i-1}} \\\nonumber
 & \approx \mathbf{\bar{H}u}+\mathbf{w},
\end{align}
where $\mathbf{\bar{H}}={{\mathbf{\hat{H}}}_{1}}{{\mathbf{F}}^{H}}$ and ${\mathbf{\hat{H}}}, {\mathbf{\hat{A}}}, {{\mathbf{\hat{H}}}_{1}}={{\mathbf{\hat{H}}}}-{{\mathbf{\hat{A}}}}$ are derived by estimated $\mathbf{\hat{h}}$. OAMP-Net is then performed to detect the transmitted OFDM symbol.

Since OAMP-Net only address real-valued variables, the complex-valued OFDM system is converted into the corresponding real-valued one before OAMP detection as in Section III. A. The equivalent real-valued version can be expressed as
\begin{align}
\mathbf{\tilde{y}}=\mathbf{\tilde{H}\tilde{u}}+\mathbf{\tilde{w}},
\end{align}
where
\[\mathbf{\tilde{y}}={{[\rm Re{{(\mathbf{\hat{y}})}^{T}}\text{ }\rm Im{{(\mathbf{\hat{y}})}^{T}}]}^{T}}, \mathbf{\tilde{u}}={{[\rm Re{{(\mathbf{u})}^{T}}\text{ }\rm Im{{(\mathbf{u})}^{T}}]}^{T}},\]
\[\mathbf{\tilde{w}}={{[\rm Re{{(\mathbf{w})}^{T}}\text{ }\rm Im{{(\mathbf{w})}^{T}}]}^{T}},
\mathbf{\tilde{H}}=\left[ \begin{matrix}
   \rm Re(\mathbf{\bar{H}}) & -\rm Im(\mathbf{\bar{H}})  \\
   \rm Im(\mathbf{\bar{H}}) & \rm Re(\mathbf{\bar{H}})  \\
\end{matrix} \right],\]
${{{{\mathbf{\tilde{u}}}}_{n}}\in \tilde{\mathcal{A}}}$,
$\tilde{\mathcal{A}}$ represents the alphabet set for the real and imaginary components of the \emph{M}-QAM signal. The model indicator $\tilde{(\cdot )}$ is dropped to obtain an uncluttered notation.

The OAMP-based detector [15] can be summarized as follows
\begin{equation}\label{1}
  {{\mathbf{r}}_{l}}={{\mathbf{\hat{u}}}_{l}}+{{\mathbf{W}}_{l}}(\mathbf{y}-\mathbf{H}{{\mathbf{\hat{u}}}_{l}}),
\end{equation}
\begin{equation}\label{2}
 \upsilon _{l}^{2}=\frac{\left\| \mathbf{y}-\mathbf{H}{{\mathbf{\hat{u}}}_{l}} \right\|-N\sigma _{w }^{2}}{tr({{\mathbf{H}}^{T}}\mathbf{H})},
\end{equation}
\begin{equation}\label{2}
\tilde{\upsilon}_{l}^{2}=(1-\beta)\upsilon _{l-1}^{2}+\beta \upsilon _{l}^{2},
\end{equation}
\begin{equation}\label{2}
{{\tau }_{l}^{2}}=\frac{1}{2N}tr({{\mathbf{B}}_{l}}\mathbf{B}_{l}^{T})\tilde{\upsilon}_{l}^{2}+\frac{1}{4N}tr({{\mathbf{W}}_{l}}\mathbf{W}_{l}^{T})\sigma _{w }^{2},
\end{equation}
 and
\begin{equation}
{{{\mathbf{{\hat{u}}}}}_{l+1}}=\mathbf{E}\{\mathbf{u}|{{\mathbf{r}}_{l}},{{\tau }_{l}}\}.
\end{equation}
In the above, $l$ indicates the index of iteration time. According to [18], the optimal matrix ${{\mathbf{W}}_{l}}$ is given by ${{\mathbf{W}}_{l}}=\frac{2N}{tr({{{\mathbf{\hat{W}}}}_{l}}\mathbf{H})}{{\mathbf{\hat{W}}}_{l}}$, where ${{\mathbf{\hat{W}}}_{l}}$ is the linear MMSE matrix, ${{\mathbf{\hat{W}}}_{l}}=\upsilon _{l}^{2}{{\mathbf{H}}^{T}}{{(\upsilon _{l}^{2}\mathbf{H}{{\mathbf{H}}^{T}}+\frac{\sigma _{w }^{2}}{2}\mathbf{I})}^{-1}}$. The matrix ${{\mathbf{B}}_{l}}$ in the algorithm is given by ${{\mathbf{B}}_{l}}=\mathbf{I}-{{\mathbf{W}}_{l}}\mathbf{H}$. Different from the OAMP algorithm in [18], the parameter update, $\upsilon _{l}^{2}$, must be smooth using a convex combination with the former value as shown in (9) to improve the robustness of the proposed algorithm. The update parameter, $\beta$, is set as 0.5. The calculation result in (10) should be non-negative. Thus, $\tau _{l}^{2}$ is replaced by $\max{(\tau _{l}^{2},\varepsilon)}$ for a small positive constant, $\varepsilon$.

From (11), ${{\mathbf{r}}_{l}}$ and $\tau _{l}^{2}$ are the prior mean and variance, respectively, which influence the accuracy of ${{\mathbf{\hat{u}}}_{l+1}}$. We use OAMP-Net to provide an appropriate step size to update ${{\mathbf{r}}_{l}}$ and $\tau _{l}^{2}$ and learn the optimal variables from a large number of data.

The structure of the OAMP-Net is illustrated in Fig. 2. The network consists of $L$ cascade layers, each with the same structure that containing the MMSE denoiser, error mean ${{\mathbf{r}}_{l}}$, error variance $\tau _{l}^{2}$, and tied weights. The input of the OAMP-Net includes the received signal, $\mathbf{y}$, and the initial value, ${{\mathbf{\hat{u}}}_{1}}\text{=}0$. The output is the estimated signal symbol, ${{\mathbf{\hat{u}}}_{L+1}}$. For the $l$-th layer of the OAMP-Net, the input includes the estimated signal, ${{\mathbf{\hat{u}}}_{l-1}}$, from the $(l-1)$-th layer and the received signal, $\mathbf{y}$.

\begin{figure}[!h]
\centering
\includegraphics[width=3in]{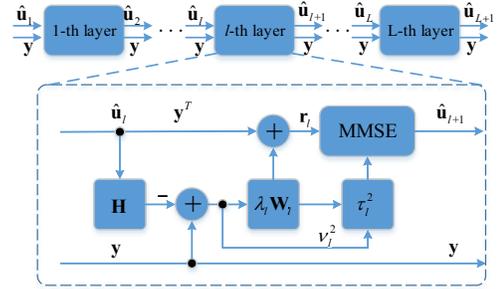}
\caption{Structure of OAMP-Net
.}
\label{fig_sim}
\end{figure}

The OAMP-Net introduces two scalar learnable parameters, $(\lambda _{l}^{{}},\gamma _{l}^{{}})$, which is different from the OAMP algorithm. Thus, (7) and (10) are transformed as
\begin{equation}
{{\mathbf{r}}_{l}}={{{\mathbf{u}}}_{l}}+{{\lambda }_{l}}{{\mathbf{W}}_{l}}(\mathbf{y}-\mathbf{H}{{{\mathbf{\hat{u}}}}_{l}}), \\
\end{equation}
\begin{equation}\label{3}
  {{\tau }^{2}}=\frac{1}{2N}tr({{\mathbf{C}}_{l}}\mathbf{C}_{l}^{T})\tilde{\upsilon} _{l}^{2}+\frac{\gamma _{l}^{2}}{4N}tr({{\mathbf{W}}_{l}}\mathbf{W}_{l}^{T})\sigma _{w }^{2},
\end{equation}
where ${{\mathbf{C}}_{l}}=\mathbf{I}-\gamma _{l}^{{}}{{\mathbf{W}}_{l}}\mathbf{H}$. If ${\lambda _{l}^{{}}=\gamma _{l}^{{}}}$, the OAMP-NET is simplified to the TISTA [14]. The transmitted symbol, $\mathbf{u}$, is obtained from the real alphabet modulation set, $\mathcal{A}=\{{{a}_{1}},...,{{a}_{m}},...,{{a}_{\sqrt{M}}}\}$ and the corresponding posterior mean estimator, $\mathbf{E}\{\mathbf{u}|{{\mathbf{r}}_{l}},{{\tau }_{l}}\}$, in (11) for each element of ${{{\mathbf{{\hat{u}}}}}_{l+1}}$ is given by
\begin{equation}\label{3}
  \mathbf{E}\{{{{{{u}^{n}}}}}|{r}_{l}^{n},{{\tau }_{l}}\}\text{=}\sum\nolimits_{{{a}_{m}\in \mathcal{A}}}{{{a}_{m}}}\frac{N({{a}_{m}};{r}_{l}^{n},\tau _{l}^{2})}{\sum\nolimits_{{{a}_{m}\in \mathcal{A}}}{N({{a}_{m}};\mathbf{r}_{l}^{n},\tau _{l}^{2})}}.
\end{equation}

There are $L$ layers in Fig. 2 and each layer contains only two adjustable variables $ (\lambda _{l}^{{}},\gamma _{l}^{{}})$. Therefore, the total number of trainable variables is equal to $2L$. Furthermore, the number of trainable variables of the OAMP-Net is independent of the number of subcarriers $N$ and is only measured by the number of layers $L$. This feature is advantageous for an OFDM system with a large number of subcarries. The trainable variables of OAMP-Net are fewer than those of ComNet [12]. The stability and speed of convergence can be improved during training.

\section{SIMULATION RESULTS}
\label{sec:typestyle}

A CP-free OFDM system with 64 subcarriers is considered in our simulation. For simplicity, each frame contains one pilot OFDM symbol and one data OFDM symbol. The wireless channel follows the wireless world initiative for new radio model (WINNER II) [19], which is also consistent with the channel models in [11] and [12]. 16QAM and 64QAM are used. CE-Net and OAMP-Net are trained by minimizing the cost between predictions and actual labels by using the adaptive moment estimator (Adam) optimizer. The learning rate is set to 0.001. For cost function, we select the $\ell_2$ loss. In CE-Net, the training and testing sets contain 3,000,000 and 1,000,000 samples, respectively. The batch size and epochs are set to 50 and 2,000, respectively.

The CE-Net is initially trained with a time-varying channel and OAMP-Net is then trained with 10,000 epochs. The OAMP-Net has 10 layers. The batch size of OAMP-Net is set to 1,000. At each epoch, the training and development sets contain 1,000 and 1,000 samples, respectively. We keep on generating the test data for the OAMP-Net until the number of bit errors exceeds 1,000. With only a single training, the DL-OAMP receiver can adapt to time-varying channels.

\begin{figure}[!h]
\centering
\includegraphics[width=2.5in]{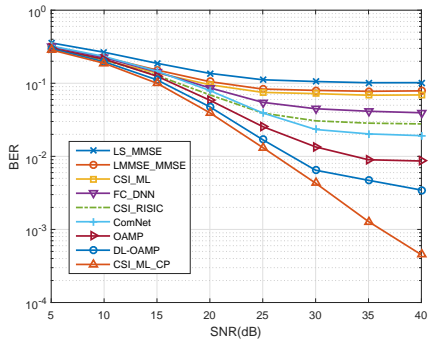}
\caption{BER curve of DL-OAMP receiver and competitive methods under the CP-free case with 64QAM.}
\label{fig_sim}
\end{figure}
The BER curve of the DL-OAMP receiver and the competitive methods in the CP-free case with 64QAM is shown in Fig. 3. \textbf{LS\_MMSE} represents traditional LS channel estimation and MMSE detection. The traditional LMMSE channel estimation and MMSE detection is denoted by \textbf{LMMSE\_MMSE}. \textbf{CSI} implies that the perfect CSI is employed for signal detection at the receiver. \textbf{ML} indicates that the maximum likelihood (ML) detector is used in the receiver. \textbf{OAMP} denotes that the CSI is obtained by CE-NET and the signal is detected by the OAMP algorithm. \textbf{CP} denotes the conventional CP-OFDM system, otherwise no CP is obtained. Classical LS\_MMSE and LMMSE\_MMSE as well as ML do not satisfy orthogonal property of subcarriers due to the effect of removing CP. Thus, all methods exhibit poor performance. On one hand, the BER of DL-OAMP is lower than those of the RISIC algorithm in [4], FC\_DNN studied by [11] and ComNet of [12]. On the other hand, the DL-OAMP can remarkably improve the performance of the OAMP algorithm. This finding indicates the superiority of OAMP-Net only by introducing a few parameters. In addition, the gap between CSI\_ML\_CP and the DL-OAMP is extremely small when the SNR is lower than 30 dB. However, the gap becomes slightly large when the SNR is higher than 30 dB.

\begin{figure}[!h]
\centering
\includegraphics[width=2.5in]{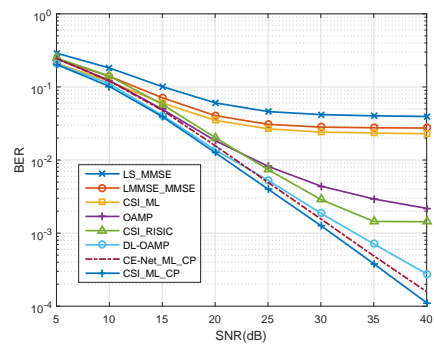}
\caption{BER curve of DL-OAMP receiver and competitive methods under the CP-free case with 16QAM.}
\label{fig_sim}
\end{figure}
Figure 4 shows BER performance for 16QAM modulation. Different from that with 64QAM, the BER of the OAMP algorithm for 16QAM is higher than that of the RISIC algorithm when the SNR is over 25 dB. However, the BER of DL-OAMP can be considerably decreased and lower than that of the RISIC algorithm. The performance improvement is because the parameters $(\lambda _{l}^{{}},\gamma _{l}^{{}})$ can be tuned in the OAMP-Net in each layer, resulting in a flexible network. In addition, the gap between DL-OAMP and CSI\_ML\_CP is small. When the SNR is blow 20 dB, the BER of DL-OAMP can approach the performance of CSI\_ML\_CP. Furthermore, DL-OAMP even outperforms the CE-Net\_ML\_CP in the case of low SNR.

In summary, the OAMP-Net exhibits excellent BER performance compared with the existing algorithms and is close to the performance limit of an OFDM system with sufficient CP for 16QAM and 64QAM.

\section{CONCLUSION}
\label{sec:majhead}

Channel estimation and signal detection are very challenging in a CP-free OFDM system. In this article, a DL based on OAMP algorithm is proposed. The channel is estimated by CE-Net, which is implemented by a two-layer neural network. For signal detection, the OAMP-Net unfolds the OAMP algorithm and adds some trainable parameters. The proposed receiver can greatly reduce the complexity of signal detection and is adjustable to time-varying channels. The simulation results demonstrate that the BER of the proposed DL-OAMP receiver is lower than those of the existing algorithms with 16QAM and other higher-order modulations.

\bibliographystyle{IEEEbib}
\bibliography{strings,refs}

\end{document}